\documentstyle[12pt]{article}

\newcommand{\beq}{\begin{equation}}
\newcommand{\beql}[1]{\begin{equation}\label{eq:#1}}
\newcommand{\eeq}{\end{equation}}
\newcommand{\be}{\begin{equation}}
\newcommand{\ee}{\end{equation}}
\newcommand{\beqn}{\begin{eqnarray}}
\newcommand{\eeqn}{\end{eqnarray}}
\newcommand{\bea}{\begin{eqnarray}}
\newcommand{\eea}{\end{eqnarray}}

\newcommand{\mx}{Matrix }

\newcommand{\IIA}{\hbox{\rm I\kern-.1emI\kern-.1emA}\ }
\newcommand{\IIB}{\hbox{\rm I\kern-.1emI\kern-.1emB}\ }
\newcommand{\athr}{A^{(3)}}\newcommand{\asix}{A^{(6)}}
\newcommand{\afour}{A^{(4)}}
\DeclareFixedFont{\xiiss}{OT1}{cmss}{m}{n}{12}
\DeclareFixedFont{\ixss}{OT1}{cmss}{m}{n}{9}
\DeclareFixedFont{\cmrnine}{OT1}{cmr}{m}{n}{9}
 
\newcommand{\CC}{\hbox{\xiiss C\kern-.45emI\kern+.15em}}
\newcommand{\RR}{\hbox{\xiiss R\kern-.5emI\kern+.15em}}
\newcommand{\HH}{\hbox{\xiiss H\kern-.85emI\kern+.5em}}
\newcommand{\ZZ}{\hbox{\xiiss Z\kern-.4emZ}}
\newcommand{\CCs}{\hbox{\ixss C\kern-.4emI}}
\newcommand{\ZZs}{\hbox{\ixss Z\kern-.4emZ}}

\begin{document}
\begin{titlepage}
\title{
	\begin{flushright}
	\begin{small}
	ILL-(TH)-97-09\\
	hep-th/9712168\\
	\end{small}
	\end{flushright}
	\vspace{1.cm}
 A Note on Six-Dimensional Gauge Theories
}

\author{
Robert G. Leigh\thanks{U.S. Department of Energy Outstanding Junior Investigator}
\thanks{e-mail: \tt rgleigh@uiuc.edu}
\ and
Moshe Rozali\thanks{e-mail: \tt rozali@hepux1.hep.uiuc.edu}\\
\\
	\small\it Department of Physics\\
	\small\it University of Illinois at Urbana-Champaign\\
	\small\it Urbana, IL 61801
}

\maketitle

\begin{abstract}
We study the new ``gauge'' theories in 5+1 dimensions, and their
non-commutative generalizations. We argue that the $\theta$-term and the
non-commutative torus parameters appear on an equal footing in the
non-critical string theories which define the gauge theories. The use of
these theories as a Matrix description of M-theory on $T^5$, as well as
a closely related realization as 5-branes in type \IIB string theory,
proves useful in studying some of  their properties.
 
\end{abstract}

\end{titlepage}

\section{Introduction}

\mx theory\cite{bfss} is an attempt at a non-perturbative formulation of
M-theory in the lightcone frame. Compactifying it on a torus $T^d$ has
been shown to be related to the large $N$ limit of Super-Yang-Mills
(SYM) theory in $d+1$ dimensions\cite{tori}. This description is
necessarily only partial for $d>3$, since the SYM theory is then not
renormalizable. The SYM theory is defined, for $d=4$ by the (2,0)
superconformal field theory in 5+1 dimensions \cite{roz, brs}, and for
$d=5$ by a non-critical string theory in 5+1 dimensions \cite{brs,
five}. The situation of compactifications to lower dimensions is unclear
for now (for recent reviews see \cite{banks,suss}).

Recent work\cite{sen,sei} on the discrete light-cone quantization (DLCQ)
of M-theory sheds light on its \mx formulation, which is the above
mentioned theories for a finite $N$.\cite{dlcq}  In particular it
clarifies the problems encountered in finding a \mx description for
compactifications of M-theory to low dimensions. It is natural, then, to
consider more general  backgrounds for \mx theory, in the cases when it
is relatively well-understood.  One  hopes that this might  uncover
phenomena relevant to the low dimensional compactifications, but in a
better controlled context.

In recent papers,\cite{connes,hullmike} it was shown that longitudinal
constant backgrounds of $\athr$ may be incorporated into \mx theory, and
lead to SYM theories on non-commutative tori.\footnote{ Other works
relating to longitudinal moduli and extended U-duality have appeared
recently\cite{uduality}.}
Indeed, the introduction of non-commutative geometry is not
surprising--it has long been suspected to appear in this context,
possibly serving as a cutoff. However, for higher dimensional tori, as
is the case for conventional SYM theories, the resulting theories are
expected to be non-renormalizable and need a UV definition.

For a large 5-torus, the compactified  SYM theory can be defined as a
particular limit of the parameter space of the non-critical string
theory. In this limit there emerges a geometrical description of the
base space, and the SYM is a low energy description in this geometrical
setting. We suggest to define the SYM theory on a large non-commutative
torus in a similar way, as a particular limit of the non-critical string
theory, where there emerges a base space which is a non-commutative 
5-torus. We refer to this scenario loosely as  ``compactifying the
non-critical string theory on the non-commutative torus''. We present a
precise definition of this theory in terms of a decoupled 5-brane theory
in what follows \footnote{ The Hilbert space of the non-critical
string theory includes states that live in the near-horizon ``throat'' \cite{ms},
but decouples from the asymptotic spacetime physics.}.
 
A different deformation of 6 dimensional gauge theories was considered
recently in Refs. \cite {wittenpq, kol}. In \cite{wittenpq} the theory
on the $(p,q)$ fivebranes of a weakly coupled type \IIB was studied. The
low energy theory is a gauge theory with a rational $\theta$-angle. An
extension to irrational values of $\theta$ was suggested in \cite{kol}.
An essential part of the analysis in \cite{kol} was the fact that the
bulk theory is not weakly coupled in any duality frame.

In this paper we take the approach advocated in \cite{incredible}. The
finite $N$ theories above are related to DLCQ of M-theory \cite{dlcq},
and have therefore a spacetime interpretation. The relation to spacetime
can be turned around to deduce statements about the  ``gauge'' theories
in 5+1 dimensions.  In order to approach these theories as \mx theories,
we are led to study compactifications of M-theory on $T^5$ with
longitudinal moduli turned on.

In this approach we reproduce the theories on the non-commutative torus
\cite{connes, hullmike}, and the ``gauge'' theories with any
$\theta$-angle \cite{wittenpq, kol}. Since there is no apparent  quantization
condition for $\theta$ in spacetime, we are led to believe that the
``gauge'' theories exist for any value of the $\theta$-angle. Moreover,
once they are defined via the non-critical string theory\cite{five},
these ``gauge'' theories are simply T-dual to the theories considered in
\cite{connes, hullmike}. This intrinsic T-duality is the U-duality of
$M(T^5)$ in the spacetime picture.

The paper is laid out as follows. We begin with a short review of
non-commutative geometry as it has appears in recent M-theory
literature. We then consider longitudinal backgrounds for $M(T^5)$, and
discuss their relation to the above mentioned gauge theories. We
describe their implications for the BPS spectrum in the low energy SYM
theory as well as in the corresponding non-critical string theory.

\section{Non-Commutative Geometry on $T^2$}

In this section, we review some key points in the recent
literature\cite{connes,hullmike} on the appearance of non-commutative
geometry in M-theory compactifications.

Consider the usual process of deriving a \mx description of $M$-theory
on a 2-torus, with no backgrounds turned on. Following Ref.~\cite{sei}, we
fix M-theory parameters, and boost the lightcone. Following this, we
must rescale all length parameters. In the resulting \IIA theory, the
length scales of the torus are small--the physics in this limit is given
by a T-dual theory on a large torus. Note that at large radius, the
light states are momentum modes, and these dominate any stringy effects. 
As the radii
get small, it is the winding modes that get light, and thus we get a
description in terms of these variables.

Now consider what happens when a longitudinal background $\athr_{12-}=\theta$
is turned on in M-theory. In the \mx description, we arrive at
\IIA theory compactified on a small torus of size $L_i$, with a $B$-field
turned on. The BPS mass spectrum becomes 
\beq
m^2=\sum_{i=1,2} \left( \frac{1}{{L_i}^2} (k_i- \theta \epsilon_{ij} w_j)^2 +
{L_i}^2 {w_i}^2 \right) + {T_{str}\over g^2_{str}}\sum \left( 
(Q_0-\theta Q_2)^2+(T_{str}L_1L_2Q_2)^2\right)
\eeq
where $Q_{0,2}$ are, respectively, D0- and D2-brane charges and
$k_i,w_i$ are momentum and winding modes. In the presence of $\theta$,
the 2-brane picks up 0-brane charge and the winding mode picks up
momentum. In the relevant limit $L_i\to0$, the winding modes are the
lightest modes. In a field theory description, one wishes to identify
these modes with momentum modes on a dual torus.

For rational $\theta$, we can describe the system by a particular sector
in a $2+1$ SYM theory; however, this description is ergodic in $\theta$.
Furthermore, for irrational $\theta$ there is no conventional Yang-Mills
description.

In the limit $L_i\to0$, the K\"{a}hler form has a vanishing imaginary
part, but retains a non-zero real part. This particular degeneration of
the torus has been shown to be best described by non-commutative
geometry. The winding modes are still described as momenta on the dual torus,
but this torus has to be taken to be non-commutative in order to reproduce the BPS formula above.

In what follows, we consider compactification on $T^5$. Clearly, if
$\athr$ is of minimal rank, say $\athr_{12-}\neq 0$, then the discussion
will follow the above--we will get SYM theories on $T^\theta\times T^3$.
There are ten such configurations possible (later, we will label these
by $\theta_{ij}$) --the generic configuration will give a general
non-commutative $T^5$. However, clearly this is only a low-energy description
and we will suggest an ultra-violet definition of this theory.

\section{Backgrounds for $M(T^5)$}

In the following we discuss gauge theories in 5+1 dimensions,
compactified on a 5-torus. As non-renormalizable field theories, they 
need to be defined in the ultraviolet. We use their description as the
low energy limit of the six-dimensional non-critical string
theories\cite{five}. We concentrate mainly on two deformations of these
theories. One of them is visible in the low energy SYM as the operator $
\theta \int Tr(F^3)$, recently discussed in Refs. \cite {wittenpq, kol}.
The other is a formulation of the gauge theory on a non-commutative
5-torus\cite{connes,hullmike}. These deformations appear very
differently in the low energy SYM theory, but are in fact closely
related in the non-critical string description.

The main tool we use to investigate the 6-dimensional string theories,
compactified on a torus, is their interpretation as \mx theories for
$M(T^5)$. The \mx description can be derived directly\cite{sen,sei},
giving the worldvolume theory of $N$ coincident NS5-branes of type \IIB
string theory. The bulk string theory has a finite string tension, and a
vanishing (asymptotic) string coupling, exactly the limit used before to
define the gauge theory\cite{five}.

The resulting theory is a non-gravitational string theory in 6
dimensions. Since it lacks an intrinsic definition, we have to resort to
indirect descriptions of it.\footnote { A \mx definition of the
non-critical string theories, and several field theory limits of them,
are discussed in Refs. \cite{ncmatrix}.} One such
description is the above embedding in type \IIB string theory. Though it
is decoupled from the bulk string theory, the intrinsic 5-brane theory
can inherit some properties of the bulk theory, provided they are
protected from quantum corrections. One such property, the T-duality
group $SO(5,5)$, serves as the U-duality group of $M(T^5)$
\cite{brs,five}. Similarly, masses of BPS saturated excitations can be
trusted \cite{five}, though their bulk interpretation (as bound states
of the NS 5-brane with various D-branes) might change in the intrinsic
theory.

We turn now to describe the \mx definition of general points of the
moduli space of M-theory on $T^5$. Moduli associated with purely
transverse directions have been described before \cite{five,bcd}, and in the
following we concentrate on "longitudinal" moduli. There are 16 such
moduli (transforming in the spinor representation of the U-duality
group):
these are the off-diagonal metric elements (angles) $g_{i-}$ and
Wilson lines (constant background values) for the 3-form $\athr_{ij-}$
and for the
6-form of M-theory, $\asix_{12345-}$. 

One can systematically derive the \mx description along the lines of
\cite {sen,sei}. \mx theory on $T^5$ is defined by  the system of $N$
0-branes in type \IIA theory with the following parameters:
\begin{eqnarray}
T_{str} &=& \frac{ R \gamma }{{\ell_p}^3}\\
{g_{str}} &=& \left(\frac { R}{ \gamma \ell_p}\right)^{3/2}\\
L_i &=& \frac{R_i}{\gamma}
\end{eqnarray}
where $\ell_p, R , R_i$ are, respectively, the 11-dimensional Planck
length, the lightcone periodicity, and the transverse $T^5$ lengths, all
in the IMF. They are kept finite as the boost parameter $\gamma$ is
taken to infinity. The $T_{str}, g_{str}, L_i$ are parameters of the
type \IIA string theory in which the 0-brane system is embedded.
In this frame, the longitudinal moduli are respectively Wilson lines
for the RR 1-form, $\theta_i$, background NS 2-forms $\theta_{ij}$
and a Wilson line for the RR 5-form, $\theta$.

Since the 5-torus is small, one needs to perform T-duality to transform
it to a 5-torus of a  finite volume. The parameters $\theta_{ij}$, appearing
in the type \IIA theory as 2-form backgrounds, would prevent, in a
conventional treatment, getting a finite volume torus using T-duality.
As explained in \cite{connes,hullmike} and reviewed above, interpreting
the parameters $\theta_{ij}$ in the framework of non-commutative
geometry allows us to describe the system as a SYM theory on a finite
size, albeit non-commutative, torus.

The resulting type \IIB theory, then, has the following parameters:
\bea
l_i &=& \frac{ {\ell_p}^3}{R R_i}\\
T_{str}&=& \frac{R \gamma}{{\ell_p}^3}\\
\tau &\equiv& \frac{i}{g_{str}}+ \frac{\theta_B}{2\pi}=
\frac{ i R R_1 R_2 R_3 R_4 R_5 }{ {\ell_p}^6 \gamma} + \theta \\
\afour_{ijkl} &=& \epsilon_{ijklm} \theta_{m}.
\eea
 
The resulting Matrix description is the system of $N$ Dirichlet 5-branes
of type \IIB theory with  infinite string coupling and tension. This is
an alternative definition of the non-critical string theory and its low
energy SYM limit. In the SYM the new moduli appear as follows:
\begin{itemize}
\item $\theta$ is the coefficient of the operator $\int Tr(F^3)$. In
fact, we have reached exactly the system used by Kol\cite{kol} to
define the generalization of the $(p,q)$ 5-brane theories of
\cite{wittenpq}.
\item $\theta_{ij}$  are the parameters of the non-commutative torus on
which the SYM is formulated \cite{connes}.
\item $\theta_i$ are  coefficients of operators $\int Tr(F_{0i})$, the
integral being over  1-cycles of the 5-torus (and time). 
\end{itemize}

This relation between the SYM couplings and the longitudinal moduli is
deduced from the well-known coupling of D-branes to RR backgrounds. As
is shown below, these parameters are related to masses of BPS saturated
states, and are therefore protected from quantum corrections.

We note that in the presence of a non-vanishing  $\theta$ one cannot, in
general, use the $SL(2,\ZZ)$ duality of type \IIB string theory to make
the bulk modes weakly interacting. For a rational $\theta$, this is
possible, yielding the $(p,q)$ theories of \cite{wittenpq}. However, in
this case the description becomes discontinuous in $\theta$, as was
noted in \cite{wittenpq}.

Thus, for a generic non-zero $\theta$, the SYM  cannot be realized as the
world-volume theory of any 5-brane in a weakly interacting string
theory. We are led then to define these theories, as in \cite{kol}, as 
{\it the world-volume theories of  D5-branes in a strongly interacting
type \IIB theory}.\footnote{ Changing in this definition the D5-brane to
a NS5-brane, or to any $(p,q)$ 5-brane, only redefines the parameter
$\theta$. The bulk theory is still strongly interacting. We fix the
relation of the gauge theory parameters to the spacetime parameters by
using D5-branes in the above definition.} It would be interesting to understand
 the detailed mechanism of decoupling from a
strongly interacting bulk theory. This could be of interest for
studying low-dimensional compactifications of \mx theory.

The situation is identical to the case considered in section 2. Indeed,
describing the configuration in  M-theory, we get $N$ M5-branes wrapping
the long cycle of a 2-torus. In the limit relevant here, the torus
complex structure has a vanishing imaginary part and a non-zero real
part. This is exactly the degeneration of the 2-torus which led
naturally to the introduction of non-commutative geometry on the base
space \cite{connes, hullmike}. In our case, lacking a clear intrinsic
description of the non-critical string theory, it is difficult to make
an analysis similar to \cite{connes}.

We end this section by commenting about the relation between the SYM on
the non-commutative torus \cite{connes, hullmike}, and the SYM theories
with $\theta$ angle. As in any string theory, the base space
geometry (commutative or not) is only a low energy artifact. In the
non-critical string theory the parameters $\theta$ and $\theta_{ij}$ are
T-dual to each other. It is only when going to a particular limit of
parameter space of the string theory that a particular parameter
manifests itself as a $\theta$ angle or as  a geometrical parameter. It is
unclear to us if the non-commutativity introduced by
these parameters is a low energy statement, or has a deeper connection to
the formulation of the  non-critical string theory itself.

\section{ Shifts in BPS charges}

In this section we discuss the BPS spectrum of the non-critical string
theories defined above, in the presence of the parameters $ \theta,
\theta_i, \theta_{ij} $. This demonstrates in detail the T-duality
transformations relating these parameters. The effect of the parameters
is to shift the charges of BPS saturated states, similar to the
well-known Witten effect \cite{witeff}. For the sake of simplicity of
notation we consider only the case of one modulus turned on at a time.
Iterating the shifts described below gives the more general situation.

The central charges of the system are all visible in the embedding of
the system in type \IIB string theory. Some of them have a description
in the SYM theory, or even as non-critical string excitations. The
central charges are the following:
\begin{itemize}
\item $m_{i}$ : In the bulk this is the winding number of the
fundamental string; it appears in the SYM theory as an electric flux.
\item $m_{ij}$: In the bulk this is a 3-brane charge in the 3-plane
transverse to $(ij)$. In the SYM theory
this is a magnetic flux in the $(ij)$ 2-plane.
\item $m_{12345}$ : In the bulk this is the NS5-brane charge. It has a
finite energy density in 5+1 dimensions, therefore it is invisible to
the SYM theory \cite{five}.
\item $N$ : the number of D5-branes in the bulk. This is the rank
of the gauge group, or its non-commutative generalization (dimension) in
the SYM theory.
\end{itemize}

All these central charges are expected to be part of the spectrum of the
non-critical string theory. The only charges that
are transparent in the non-critical string frame are:
\begin{itemize}
\item $k_i$ : This is the momentum conjugate to translations on the
5-torus, from each point of view (SYM, non-critical string, type \IIB).
\item $w_i$ : This is the D1-brane number in the bulk. In the 5-brane
theory it is a winding number of the non-critical string, and appears in
the SYM as the instanton number, $I_{jklm}$, in the 4-plane transverse to $i$.
\end{itemize}

Turning on a single modulus corresponds to moving in an $SL(2,\RR)$
subgroup of the moduli space of $M(T^5\times S^1_{LC})$. One then arranges these
central charges into representations of the particular $SL(2,\RR)$
involved, and borrows the results from \cite {witeff}. The ``magnetic''
charge is left unchanged, whereas the ``electric'' charge is shifted.

One needs to decide which is the magnetic, and which is the electric charge.
The former is
the charge  which, in the limit involved, appears as a solitonic,
topologically quantized, charge. In order to make this
identification we use the spacetime interpretation of the charges involved,
via \mx theory.

Using this assumption we get the following results:

\begin{itemize}
\item When turning on the backgrounds $\theta_{ij}$, we get the
following shifts in the charges which appear in the BPS mass formula 
(we use $i,j,k,l,m$ as a set of cyclic
indices of the 5-torus coordinates):
	\begin{itemize}
	\item $N \rightarrow N- \theta_{ij} m_{ij}$. 
	\item $ m_i \rightarrow m_i - \theta_{ij} k_j$.
	\end {itemize}  
These effects were observed in \cite{connes}, and are explained by the
non-commutative geometry of the base space. We note that these shifts
are not expected to be seen in a conventional gauge theory. Turning on
$\theta$-like terms indeed cause  mixing between  charges in the gauge
theory. However, the charges that are ``fundamental'' shift, while the
``solitonic'' charges stay fixed. From the above formulas we see that
the fixed charges are $k_i, m_{ij}$, while the charges that shift are
$N, m_i$. In that sense we are working in a ``dual'' formulation of the
theory where the ``fundamental'' charges are $N, m_i$, which are
solitonic in the usual semi-classical formulation of the gauge theory.

The relation to spacetime, as well as the intrinsic T-duality predicts
one more shift:
	\begin{itemize}
	\item $m_{kl} \rightarrow m_{kl} - \theta_{ij}I_{ijkl} = 
		m_{kl} - \epsilon_{ijklm} \theta_{ij} w_m $
	\end {itemize}  
     
\item When turning on the backgrounds $\theta_i$ one gets:
     
	\begin{itemize}
	\item $N \rightarrow N - \theta_i m_i$.
	\item $m_{ij}\rightarrow m_{ij} - \theta_i k_j$.
	\item $m_{12345} \rightarrow m_{12345}- \theta_i w_i$.
	\end{itemize}  

As above, these are not the usual shifts in the gauge theory. For
example the usual treatment of the term $ \theta_i \int Tr(F_{0i})$
would shift the electric fluxes $m_i$, while keeping $N$ fixed.

\item Finally, when turning on $\theta$, one gets:
     
	\begin{itemize}
	\item $N \rightarrow N -  \theta m_{12345}$.
	\item $m_i \rightarrow m_i - \theta w_i $.
	\end{itemize}
The first shift is invisible to the SYM theory. The second is an
electric flux carried by instantons in the presence of the $ \theta \int
Tr(F^3) $ term.
     
\end{itemize} 

We may summarize all of these shifts by arranging them into representations
of the T-duality group, $SO(5,5)$. We have the singlet $N$, a spinor
$M=(m_i,m_{ij},m_{12345})$, a vector $\Phi=(w_i,k_i)$, and an
(anti)spinor $\bar\Theta=(\theta_i,\theta_{ij},\theta)$. The shifts are then
the $SO(5,5)$ covariant expressions
\beqn
M&\to& M-\Phi\cdot\bar\Theta\\
N&\to& N-\bar\Theta M
\eeqn

It should be possible to realize $\theta, \theta_i,\theta_{ij}$ 
as parameters of the intrinsic 5-brane theory. A hint, and a
constraint on any such formulation are the shifts above. We note
that the BPS masses vary continously with $\Theta$.

To summarize, we have discussed an ultra-violet definition of gauge theories in
$5+1$-dimensions. These include gauge theories on non-commutative tori,
as well as those with an arbitrary $\theta$-angle. These theories can be
defined in terms of the non-critical string theories in six dimensions. An 
intrinsic definition of these theories is lacking, but we are able to define
them formally in terms of their embedding in a particular background of type \IIB 
string theory. This embedding enables us to identify the BPS spectra of
these theories.

Work supported in part by DOE grant DE-FG02-91ER40677.

\bibliographystyle{unsrt}

\def\npb#1#2#3{Nucl. Phys. {\bf B#1} (#2) #3}
\def\plb#1#2#3{Phys. Lett. {\bf #1B} (#2) #3}
\def\prd#1#2#3{Phys. Rev. {\bf D#1} (#2) #3}
\def\prl#1#2#3{Phys. Rev. Lett. {\bf #1} (#2) #3}
\def\mpl#1#2#3{Mod. Phys. Lett. {\bf A#1} (#2) #3}
\def\hepth#1#2#3#4#5#6#7{hep-th/#1#2#3#4#5#6#7}

\end{document}